# High-current superconductor transport critical-current measurement option for the Quantum Design Physical Property Measurement System

N. M. Strickland,[1] A. Choquette[1,2], E. F. Talantsev[1,3,4] and S. C. Wimbush[1,1)]

[1]*Robinson Research Institute, Victoria University of Wellington, PO Box 600, Wellington 6140, New Zealand*

[2]*Institut quantique and Département de physique, Université de Sherbrooke, Sherbrooke, Québec J1K 2R1, Canada*

[3]*Present address: M. N. Mikheev Institute of Metal Physics, Ural Branch, Russian Academy of Sciences, 18 S. Kovalevskoy St., Ekaterinburg 620108, Russia*

[4]*Present address: NANOTECH Centre, Ural Federal University, 19 Mira St., Ekaterinburg, 620002, Russia*

We report on the design and operation of a transport critical-current measurement option for superconductors based on the widely used Physical Property Measurement System from Quantum Design. The system is capable of supplying transport currents up to 30 A while maintaining a sample temperature of $2.0 \pm 0.1$ K, and currents up to 200 A at higher sample temperatures.

Transport critical current characterization is ubiquitous in fundamental [1-3] and applied [4-7] superconductivity, especially for the case of high-temperature superconducting (HTS) wires and tapes which operate over an extended range of temperatures and magnetic fields. The key figures of merit for wire measurement systems [8-12] are the base temperature to which the sample can be cooled, the transport current that can be supplied to the sample at that temperature without significant sample heating, and the magnetic field that can be applied to the sample during measurement. In the early stages of second-generation (2G) HTS wire development [13], it was common to characterize samples only at liquid nitrogen temperature (77 K), but it was subsequently found that these results do not correlate well with the low-temperature behavior and consequently extended-temperature measurements are essential [14].

We have previously developed a measurement system based on sample cooling by a closed-cycle flow of helium gas cooled by heat exchangers on a cryocooler that allows us to cool samples to 20 K while supplying transport currents up to 875 A under magnetic fields up to 8 T [9]. However both applied and fundamental physics would benefit from access to the

---

[1)] Author to whom correspondence should be addressed. Electronic mail: stuart.wimbush@vuw.ac.nz.

temperature range 2 – 20 K as well as the higher temperatures we currently access. Some of the more recently discovered superconductors, such as magnesium diboride [15] and the iron-based superconductors [16], have intermediate-range superconducting transition temperatures (30 – 50 K). These would typically operate at or below 20 K in applications, so it becomes important to be able to characterize wires in the liquid helium range while retaining the capability of supplying transport currents of the order of tens of amperes. There is interest also from a fundamental perspective to measure the low-temperature self-field critical current density, $J_c^{sf}(T)$. As recently reported [17], the ground-state London penetration depth, $\lambda(0)$, and superconducting energy gap, $\Delta(0)$, can be extracted from the temperature dependence of $J_c^{sf}(T)$ to sufficiently low temperatures.

The traditional approach to achieving low temperatures of immersing the superconductor in liquid helium is convenient but expensive and does not easily allow for the sample temperature to be varied. Commonly a stream of helium gas boil-off from a liquid helium bath is used to provide variable cooling to achieve a wide range of temperatures. This is the method used, for example, in the Quantum Design Physical Property Measurement System (PPMS).

As many laboratories worldwide already own and operate a PPMS, we chose to develop a new high-current transport critical current measurement option based on this system as an add-on that can be easily adopted at relatively low cost. The PPMS has built-in capability to apply an axial magnetic field (up to 9 T in the case of our model), to cool the sample space down to 1.9 K and to supply transport currents up to 0.1 A dc (as standard) or up to 2 A ac (using the ac transport option). This magnitude of current is too low to effectively study practical HTS wires as current transport bridges cannot be made sufficiently narrow without introducing spurious size-related effects [18]. Accordingly, we aimed to make use of the built-in capability of the PPMS in terms of magnetic field and temperature control, while supplying our own high dc transport current to a custom-designed sample mounting rod through an external instrumental setup based upon our existing system.

The basic design of the sample mounting rod and high current leads was similar to that of our previous system [9] where the sample was cooled by circulating cold helium gas. This is an effective method for cooling high-current leads since the there is much cooling power remaining in the gas after cooling the sample. The PPMS does not have an option to circulate the cooling gas so we use static cold helium gas instead. This limits current lead cooling and so some effort was made to reduce heat leaks from the PPMS top flange to the end of the sample and of the rod. One aspect of this was the incorporation of a series of baffles along the length of the rod that both prevent radiative heating and help to stabilize the temperature profile of the static gas. We sized the rod such that the sample was positioned about 5 cm above the bottom of the cold insert of the PPMS when inserted, in a similar position within the homogeneous magnetic field region as for other PPMS options such as the horizontal rotator or vibrating sample magnetometer.

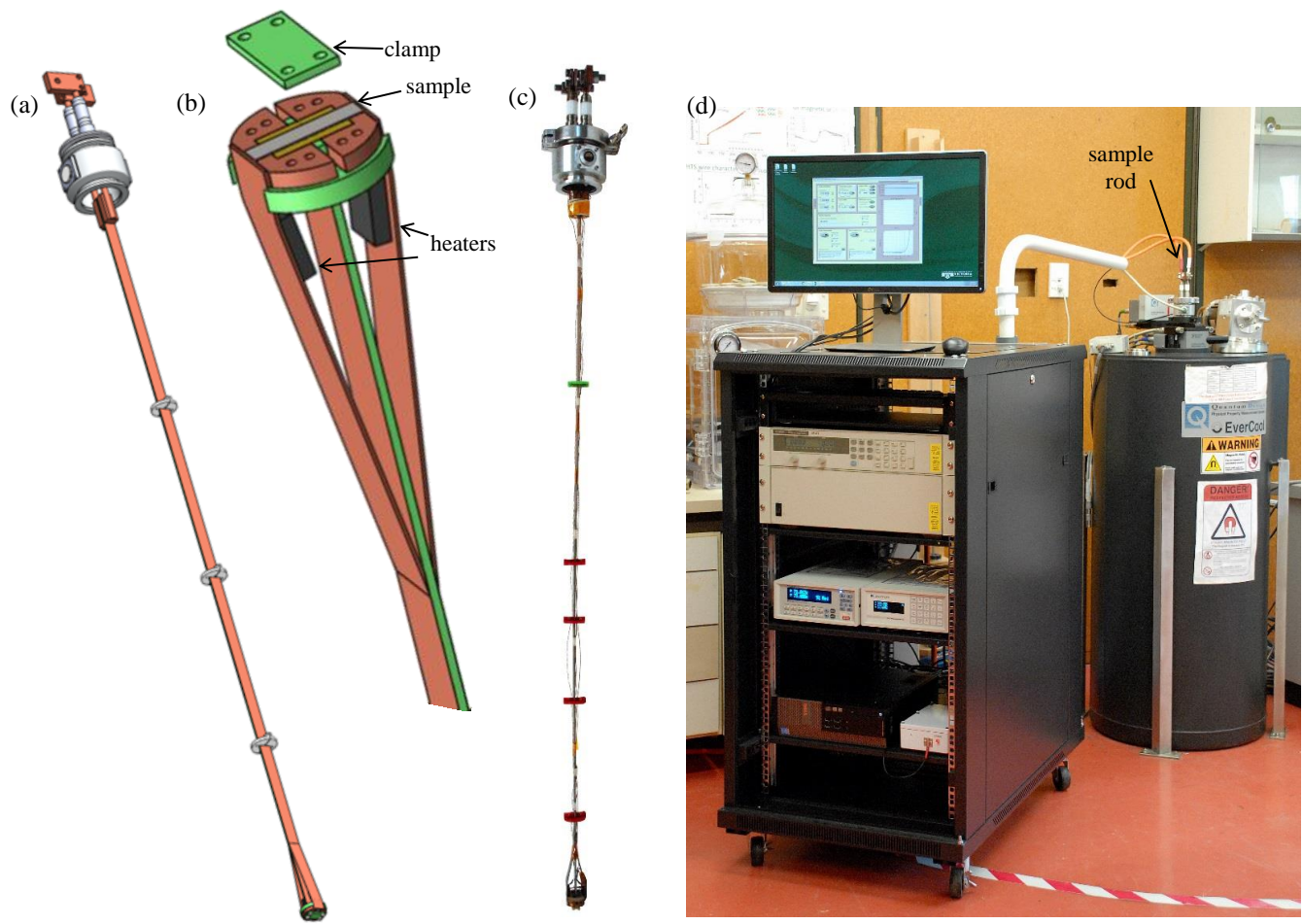

Fig. 1. Schematic diagrams and photographs of the developed measurement option: (a) custom sample rod and (b) detailed view of the sample mounting region showing the split current contacts, sample clamp, sample heaters and sample temperature sensor; (c) photograph of the constructed sample rod; (d) photograph of the external instrumentation setup interfaced to the PPMS system.

A second aspect of reducing heat leaks was the construction of sample rod and current leads. The sample rod (Fig. 1a,b) was constructed from a G10 strip 850 mm long, 12 mm wide and 1.6 mm thick, plated with 30 µm of copper on each side. This copper plating served primarily as a substrate to which to solder lengths of Sumitomo DI-BSCCO HTS wire which effectively carry the current through their bulk silver matrix at the hot end of the rod and through the superconductor at the cold end. The selection of these conductors is a balance between the heat conduction from room temperature down the rod and the Joule heating of the normal-state part of the current leads when the current is flowing. By isolating the sample end of the rod from the Joule heating through the use of a superconducting current lead, we can delay the heat transfer to the sample until after the measurement is completed. Three Lake Shore DT-470 silicon diode temperature sensors were positioned along the rod to monitor its temperature profile and prevent overheating.

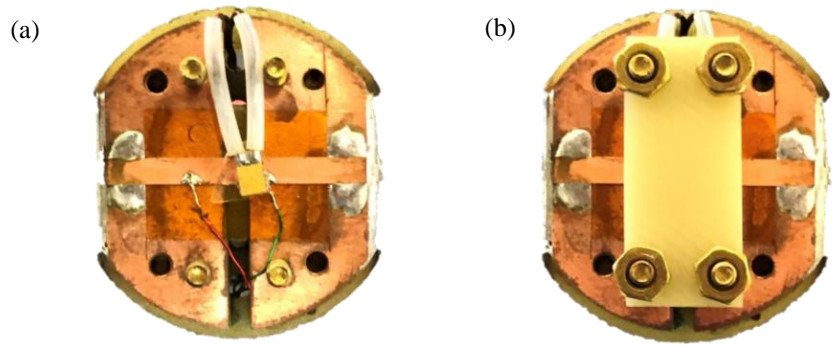

Fig. 2. Photograph of (a) a mounted 2G wire sample soldered to the holder, with voltage taps soldered to the lower edge and Cernox temperature sensor glued to a sapphire plate, and (b) the clamp in place to secure the temperature sensor against the sample.

The sample is mounted horizontally (within the vertical field of the PPMS) on the end of the sample rod on a split copper plate separated by the continuation of the G10 strip and held together by an insulating G10 block (Fig. 2). The sample is soldered to the current leads using a low temperature In-Bi solder at either end. Voltage leads running along the sample rod are passed through the central gap in the mount and soldered directly to the sample. The sample used here to demonstrate the instrument was a 2 mm wide copper-plated SuperPower tape type SCS2030-AP. The gap between the two current leads is small – only the thickness of the G10 strip – so a piece of insulating Kapton tape was inserted under the central part of the sample to better define the distance between current contacts and ensure adequate current transfer into the superconductor.

The sample temperature is monitored using a Cernox temperature sensor that was glued to a 3 mm × 2 mm × 0.2 mm sapphire plate and clamped in contact with the sample using a clamping plate made of G10 that also serves to restrain the sample against the lateral Lorentz force. Application of Apiezon N low-temperature grease ensured good thermal contact between the sapphire plate and the sample. The sample temperature is controlled using two resistive heaters attached to equivalent positions on each of the current leads running to the sample, as shown on Fig. 1a. The transport current setup, electronics and control software were modified versions of those reported by us previously [9].

The PPMS "Vent and Seal" operation enables us to seal the sample space with an atmospheric pressure of He vapour after inserting the sample rod at room temperature and performing the usual purging procedure. This pressure reduces to around 50 Torr upon cooling the PPMS insert to 4 K. This was found to be a sufficient heat transfer gas pressure to effectively cool the sample mounted on the sample rod to a temperature about 0.1 K higher than the set temperature of the PPMS insert. We use an external Lake Shore Model 335 Temperature Controller to control the sample temperature with a stability better than 10 mK using the heaters built into the sample rod.

Examples of the actual measurement data underlying this analysis are shown in Fig. 4a and replotted on a logarithmic scale after subtraction of the linear component to highlight the noise level on Fig. 4b. This data was obtained on cooling the sample in zero applied field to successively lower temperatures until the current capacity of the sample rod was exceeded above 200 A. An electric field criterion of $E = 1$ µV/cm to define $I_c$ corresponds to a voltage level of 0.5 µV for the 5 mm voltage tap spacing, which lies clearly above the noise.

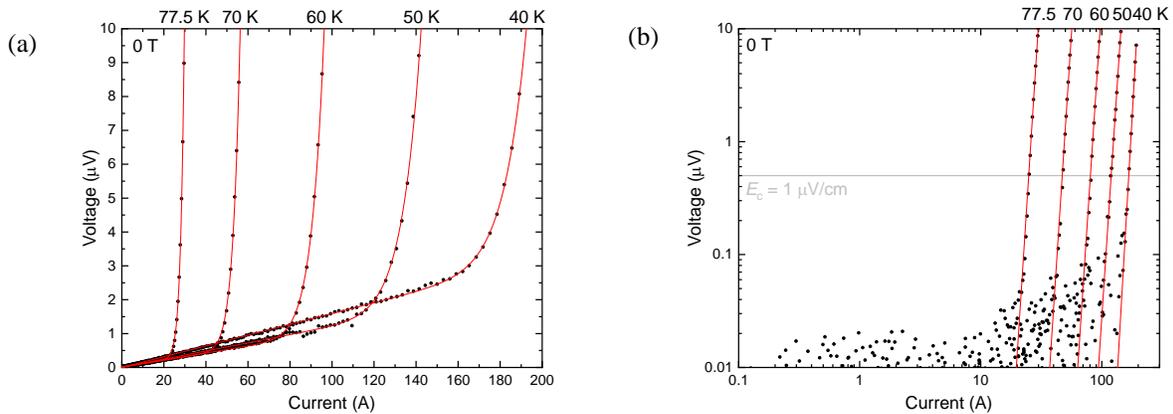

Fig. 4. Example *IV* curves up to 200 A (a) as measured, and (b) after subtraction of the linear component resulting from incomplete current transfer into the sample (plotted on a double logarithmic scale to highlight the noise level). The criterion for determination of $I_c$ is marked.

To demonstrate measurement at low temperature, a magnetic field of 9 T was applied to suppress $I_c$ sufficiently to be measurable, and the *IV* curve shown on Fig. 5 was obtained at 4.2 K. This demonstrates measurement at the extremes of the system capability. The sample temperature rise during this measurement was 0.2 K, at the borderline of acceptability. At lower temperatures, where $I_c$ is higher, the sample temperature rise at these currents was unacceptable.

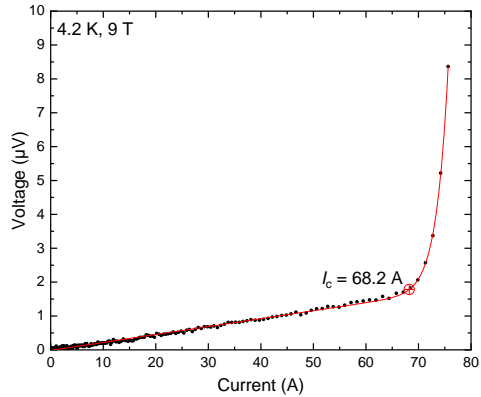

Fig. 5. *IV* curve measurement obtained at 4.2 K, 9 T, showing a sample $I_c$ of 68.2 A. The sample temperature at $I_c$ reached 4.4 K.

Finally we illustrate in detail the sample temperature stability under the transport current ramp. The instantaneous sample temperature is recorded at each current setpoint during the ramp. In Fig. 6a, we show the sample temperature variation during a measurement sweep for sample temperatures ranging from 2 K to 9 K. As the sample temperature setpoint reaches the lower limit of the PPMS capability and no excess cooling power is available to maintain the temperature against sources of heating, we see that the sample begins to rise in temperature earlier in the current sweep. This sets a practical limit on the current able to be supplied by the system for a given acceptable sample temperature rise, such as (for example) 0.1 K. On Fig. 6b we show for each temperature setpoint the current at which the rise in sample temperature exceeds 0.05 K, 0.1 K and 0.25 K. From this we see for example that we can maintain a sample temperature of 2.0 K within a tolerance of 0.1 K up to a current of 30 A, rising to 40 A at 4.2 K.

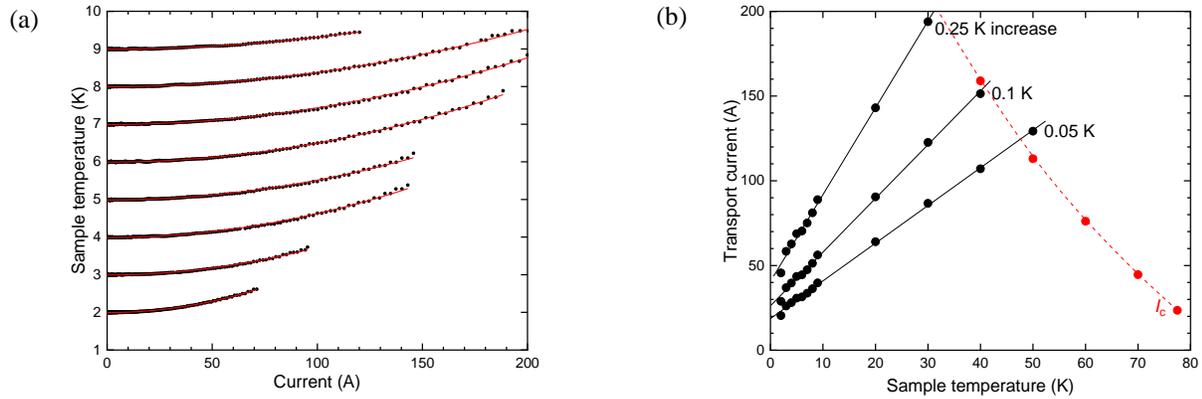

Fig. 6. (a) Sample temperature during a transport current ramp at low temperatures. (b) Transport current at which sample temperature rise exceeds various limits for different sample temperatures.

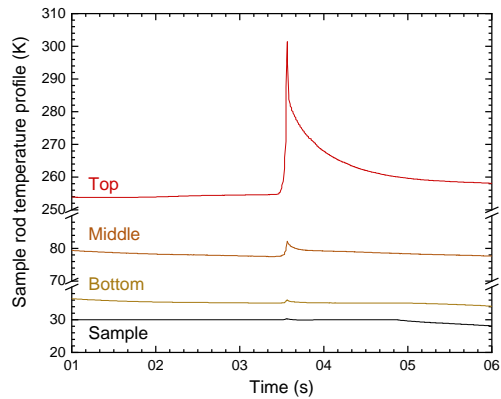

Fig. 7. Sample rod temperature profile variation during a current ramp to 200 A with the sample held at 30 K.

At high sample temperatures, the current limitation is provided not by sample heating but by overheating of the sample rod itself. This is exemplified in Fig. 7, which plots the variation in the sample rod temperature profile during a measurement of the sample at 30 K, where its $I_c$ exceeds 200 A. Although the temperature rise on the sample is limited to 0.3 K, the temperature rise along the rod grows from 1 K at the lower end to 5 K around the middle, and almost 50 K at the top end. It is this overheating of the upper end of the rod that prevents the attainment of higher currents. This could potentially be addressed by strengthening the current capacity of the rod by the addition of further copper to the upper end, with only a small impact on the base temperature.

To demonstrate the utility and operational performance of the system we show in Fig. 3a the temperature dependence of the self-field critical current, $I_c^{sf}(T)$, measured on a second SuperPower tape with an HTS layer thickness of $2b = 1$ μm and a reduced with defined by a lithographically patterned bridge of width $2a = 0.12$ mm. We have shown [17] that for thin weak-link-free type-II superconductors, the self-field critical current density obeys the following equation:

$$J_c^{sf}(T) = \frac{\Phi_0}{4\pi\mu_0} \cdot \frac{\ln \kappa_c + 0.5}{\lambda_{ab}^3(T)} \tag{1}$$

where $\mu_0 = 4\pi \times 10^{-7}$ H/m is the permeability of free space, $\Phi_0 = 2.068 \times 10^{-15}$ Wb is the magnetic flux quantum, $\kappa_c = \lambda_{ab}/\xi_{ab} = 95$ for YBa$_2$Cu$_3$O$_7$ [19] is the Ginsburg-Landau parameter and $\lambda_{ab}$ is the in-plane London penetration depth. Thus, the value of $\lambda_{ab}(T)$ may be deduced from the above equation, all other parameters being known, and these values are also shown in Fig. 3a. The theoretical weak-coupling clean-limit $d$-wave $\lambda_{ab}(T)$ curve [20] and the corresponding $I_c^{sf}(T)$ calculated from Eq. 1 are shown as dashed lines. Agreement between the experimental and theoretical results is generally very good. Some deviation in the shape of the theoretical curve from the experimental data results from the fact that real superconductors do not exactly obey the theoretically simplest case of the clean limit. The green data point in Fig. 3a is a reference value for the ground state in-plane London penetration depth, $\lambda_{ab}(0) = 125$ nm, measured on a YBa$_2$Cu$_3$O$_7$ single crystal by muon-spin rotation [21]. The correspondence between the reference value and the extrapolated value of the scaled $d$-wave curve is also very good, taking into account that 2G wires have a large density of defects that cause an increase in the penetration depth of the material, pushing it out of the clean limit.

Highlighting the true utility of the new measurement capability, we deduce on Fig. 3b the ground-state London penetration depth, $\lambda_{ab}(0)$, and the superconducting energy gap, $\Delta(0)$, of the SuperPower sample from a fit of $\lambda_{ab}(T)$ at low temperature to the low-temperature asymptote of the Bardeen-Cooper-Schrieffer (BCS) theory [22]:

$$\lambda(T) = \lambda(0)\left(1 - \sqrt{2}\frac{k_B T}{\Delta_m(0)}\right)^{-1/2} \tag{2}$$

where $k_B = 1.38 \times 10^{-23}$ J/K is the Boltzmann constant and $\Delta_m(0)$ is the maximum amplitude of the ground state $k$-dependent $d$-wave superconducting gap, $\Delta(0) = \Delta_m(0)\cos(2\theta)$. The resulting values for $\lambda_{ab}(0) = 133.2 \pm 0.1$ nm and $\Delta_m(0) = 16.1 \pm 0.3$ meV are in excellent agreement with expectation, especially the value of the superconducting energy gap, which yields the BCS ratio $2\Delta_m(0)/k_B T_c = 4.24 \pm 0.16$ for a transition temperature $T_c = 88$ K, which is remarkably consistent with the $d$-wave BCS weak-coupling limit of 4.28 [20]. These results convincingly demonstrate the utility of the system in opening up new regimes of measurement and analysis at low temperatures and high currents.

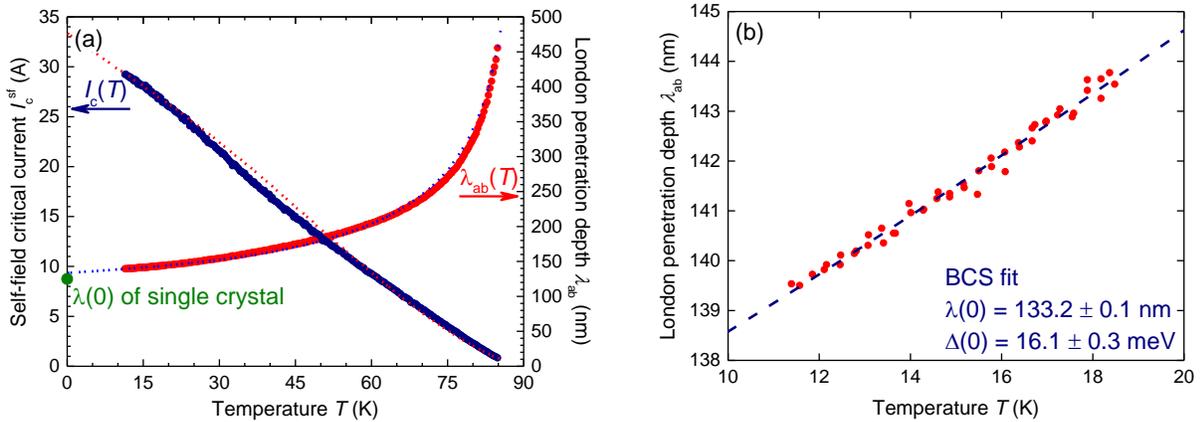

Fig. 3. Temperature dependence of the self-field critical current $I_c^{sf}(T)$ and deduced London penetration depth $\lambda_{ab}(T)$ for SuperPower 2G HTS wire. (a) Data across the full temperature range below $T_c$. Dotted lines superimposed on the data are theoretical curves for a weak-coupling clean $d$-wave superconductor. The green data point is the reported $\lambda_{ab}(0)$ value for a $YBa_2Cu_3O_7$ single crystal [21]. (b) Low-temperature BCS fit of deduced London penetration depth $\lambda_{ab}(T)$ providing superconducting parameters $\lambda(0)$ and $\Delta(0)$.

In summary, we have designed and constructed a high-current transport critical current measurement option for superconductors building on the platform of a Quantum Design PPMS. The system is capable of supplying up to 30 A of dc current to a sample held at 2 K, or 45 A to a sample held at 5 K, without appreciable sample heating. We have demonstrated the successful operation of the system through measurement of the temperature-dependent critical current of a commercial 2G HTS wire to low temperatures, and determination of various fundamental ground state parameters associated with the superconductor.

The visit of A. C. to the Robinson Research Institute was funded by an International Internship from the Université de Sherbrooke.